\documentclass
[preprint,prl,twocolumn,amsfonts,amssymb,byrevtex,tightenlines,10pt,nobalancelastpage]{revtex4}%
\usepackage{amsfonts}
\usepackage{amsmath}
\usepackage{amssymb}
\usepackage{graphicx}%
\providecommand{\U}[1]{\protect\rule{.1in}{.1in}}

\begin{document}
\preprint{ }
\title{Magnetic Field Induced Superconductivity in Out-of-Equilibrium Nanowires}
\author{Yu Chen, S. Snyder, and A. M. Goldman}
\affiliation{School of Physics and Astronomy, University of Minnesota, Minneapolis, MN
55455, USA}
\pacs{PACS number}

\begin{abstract}
Four-terminal resistance measurements have been carried out on Zn nanowires
formed using electron-beam lithography. When driven resistive by current,
these wires re-enter the superconducting state upon application of small
magnetic fields. The data are qualitatively different from those of previous
experiments on superconducting nanowires, which revealed either negative
magnetoresistance near T$_{c}$ or magnetic field enhanced critical currents.
We suggest that our observations are associated with the damping of phase slip
processes by the enhancement of dissipation by the quasiparticle conductance
channel resulting from the application of a magnetic field.

\end{abstract}
\volumeyear{2009}
\volumenumber{number}
\issuenumber{number}
\eid{identifier}
\date[Date text]{date}
\received[Received text]{date}

\revised[Revised text]{date}

\accepted[Accepted text]{date}

\published[Published text]{date}

\startpage{1}
\endpage{4}
\maketitle

Systems of reduced dimension are frequently governed by quantum behavior not
found in bulk materials. The Tomonaga-Luttinger Liquid phenomenon in one
dimension is a striking example \cite{Glazman}. Quasi-one dimensional
superconductors, whose widths and thicknesses are larger than the Fermi
wavelength but already smaller than the superconducting coherence length, are
not exceptions. There has been on-going interest in problems such as the
crossover between thermal and quantum phase slip processes \cite{Zaikin}, and
the superconductor-insulator transition controlled by either total wire
resistance or resistance per unit length in such systems \cite{Bezryadin}. In
addition, attention has been focused on the enhancement of superconductivity
by an applied magnetic field, which has been reported as a negative
magnetoresistance in some cases \cite{Xiong,Santhanam} or an enhancement of
the critical current in others \cite{Rogachev,Vodolazov}. Recently a
phenomenon called the \textquotedblleft antiproximity effect\textquotedblright%
\ has been reported \cite{Tian}. In this work, superconducting nanowires were
prepared using an electrochemical technique and connected in a two-terminal
arrangement to electrodes with higher transition temperatures. The wires,
because of their confined geometry, have a higher critical magnetic field than
the electrodes. In contrast with the usual proximity effect, at certain
temperatures the wires were found to re-enter the superconducting state when
the electrodes are driven normal by a magnetic field. The present work was
motivated by the goal of seeing whether this phenomenon would occur in wire
configurations prepared using a top-down lithographic technique rather than a
bottom-up electrochemical technique and whether the effect could be observed
in a four-terminal planar configuration. In this Letter we report a different
phenomenon, reentrant superconductivity resulting from the application of
small magnetic fields to wires driven out of equilibrium and into a resistive
state by externally supplied currents.

Standard four-terminal configurations of an 80nm wide Zn wire with 1$\mu$m
wide Zn electrodes, 1.5$\mu$m apart, as shown in Fig.1(a), were prepared using
electron-beam lithography. The 150 nm thick Zn films for the wires and
electrodes were deposited in a single step at a rate of 6\AA /sec onto
SiO$_{2}$ substrates held at 77 K. The system pressure was around
$1\times10^{-7}$ Torr during deposition and the starting material was of
99.9999\% purity. The relatively small size of the Zn grains formed under
these conditions ensured continuity of the resultant wires. The issue of the
fragile nature of the liftoff process for these samples was circumvented by
utilizing a bilayer of PMMA 495K A4/950K C2 as the resist. In order to
minimize surface oxidation, the wires were immediately transferred after
liftoff into a high vacuum and low temperature environment, a Quantum Design
Physical Properties Measurement System (PPMS) equipped with a $^{3}$He insert.%

\begin{figure}
[ptb]
\begin{center}
\includegraphics[
height=2.0038in,
width=3.3918in
]%
{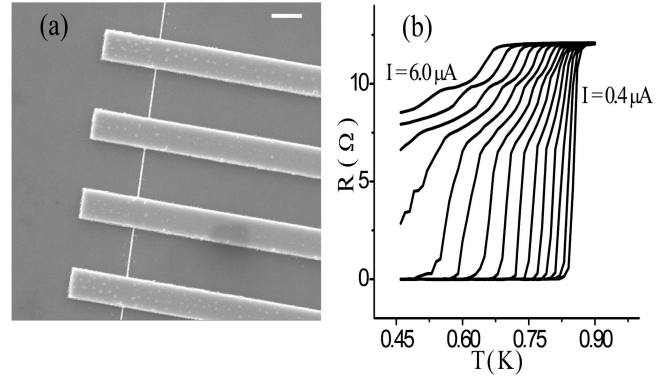}%
\caption{a) Scanning Electron Microscope (SEM) image of the sample, the white
scale bar is 1$\mu m$ long. b) Temperature dependence of the wire resistance,
at H = 0 Oe, with current ranging from 0.4$\mu A$ to 6 $\mu A$, every 0.4 $\mu
A$.}%
\end{center}
\end{figure}

In the low current limit, the temperature dependence of the wire resistance
was quite conventional. As shown in Fig.1(b), the resistance dropped to zero
at $T_{c}\sim0.85$ K with a width of a few tens of mK. We estimated the
zero-temperature coherence length, $\xi(0)\sim\left(  \xi_{0}l_{e}\right)
^{1/2},$ to be around 2100 \AA , where $\xi_{0}$ is the BCS coherence length,
and $l_{e}$ is the mean free path. Here we used the same approach as that
employed in the antiproximity effect work to obtain $l_{e}$ from the product
$\rho_{_{Zn}}l_{e}=2.2\times10^{-11}\Omega\cdot cm^{2}$ at 4.2 K \cite{Schulz}.

As the applied current increased, the onset temperature decreased and the
transition broadened to several hundreds of mK in the high current limit.
Accompanying the broadened transition was a shoulder-like structure, which
separated the transition into two parts.%

\begin{figure}
[ptb]
\begin{center}
\includegraphics[
height=1.5973in,
width=3.3918in
]%
{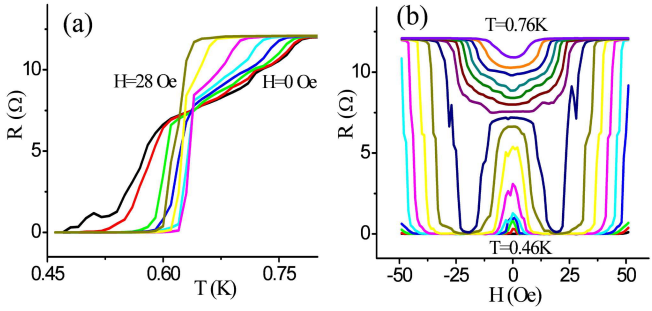}%
\caption{(Color online) a) Temperature dependence of the wire resistance, at I
= 4.4 $\mu A$, with varying applied magnetic fields from 0 Oe to 28 Oe, every
4 Oe. b) Magnetic field dependence of the wire resistance, at I = 4.4$\mu A$,
with temperatures ranging from 0.46K to 0.76K, every 0.02K. }%
\end{center}
\end{figure}

As shown in Fig. 2(a), the higher resistance part of the transition moved to
lower temperatures with increasing magnetic field. The lower resistance part
exhibited a different behavior, moving to higher temperatures with increasing
field. Also the temperature at which the wire resistance vanished increased.
As a result, the transition became sharper with increasing field. However this
eventually stopped and the transition onset temperature as well as the
temperature at which the resistance vanished, both moved together towards
lower values upon increasing the field. The direct consequence is that there
is a magnetic field induced re-entrance into the superconducting state over
the range of temperatures corresponding to the lower part of the zero field
transition as illustrated in Fig. 2(b). Over this range of temperatures, the
wire is made resistive by applying a high current. Magnetic field drives the
wire superconducting until the field is strong enough to destroy the amplitude
of the order parameter. In the higher temperature regime, a magnetic field
only suppresses superconductivity.%

\begin{figure}
[ptb]
\begin{center}
\includegraphics[
height=4.7971in,
width=2.0081in
]%
{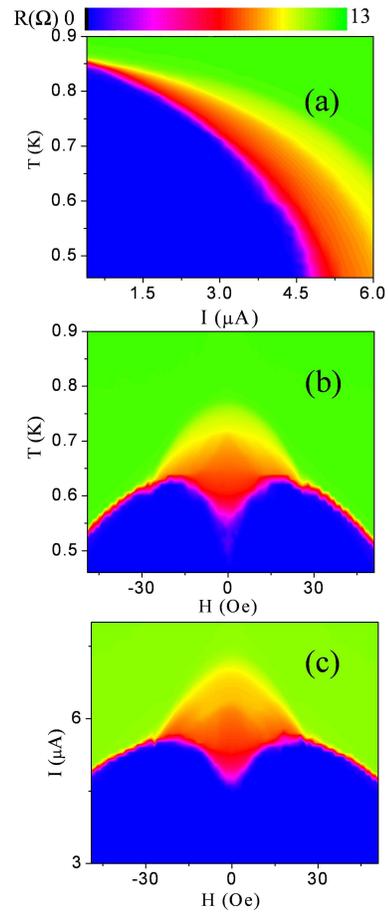}%
\caption{Color contour plot of the wire resistance as a function of a)
temperature and applied current, at H = 0Oe. b) temperature and applied
magnetic field, at I = 4.4$\mu A$. c) applied current and magnetic field, at T
= 0.46K.( The color scale bar represents the resistance of the wire, where
blue corresponds to zero resistance and green to the normal state
resistance.)}%
\end{center}
\end{figure}

Plotting the data as color maps, as shown in Fig.3, permits us to identify
three states of the wire, the normal state(green), the superconducting
state(blue) and the resistive state(colors between these two). The latter is
the transition regime. Increasing the current not only moves the transition
regime to lower temperatures but also greatly broadens it (Fig.3(a)). When the
magnetic field is turned on, shown in Fig.3(b), it gradually narrows this
regime by pulling down the boundary between the normal state and the resistive
state while pulling up the boundary between the resistive state and the
superconducting state. This gives rise to a magnetic field induced re-entrance
into the superconducting state at the bottom part of the transition regime.
The current dependence of the wire resistance is similar to its temperature
dependence, shown in Fig.3(c). In zero field, the superconductor-normal metal
transition with current is broad and exhibits a shoulder-like structure. A
relatively small applied magnetic field moves the current at the threshold for
resistance to higher values, and the current at which the normal resistance is
attained to lower values. As a consequence, superconductivity reappears in
weak magnetic fields, at currents slightly higher than the critical current at
which zero resistance disappears in zero field. This enhancement disappears a
higher fields, or at currents above the shoulder. This is different from the
anti-proximity effect, in which the wire switches abruptly from the normal to
the superconducting state when the magnetic field reaches the critical field
of the bulk electrodes. The re-entrance into the superconducting state here is
a smoother and broader transition from the resistive state. In addition, the
magnetic field needed is much weaker than the critical field of the bulk
electrodes and its value is a function of temperature and current.

Before considering possible physical mechanisms for this \textquotedblleft
magnetic-field enhanced superconductivity,\textquotedblright\ we need to rule
out several other phenomena, which might produce similar results. One
possibility is that the wires are heated by the currents and the effect of the
magnetic field is to enhance their thermal conductivity by increasing the
quasiparticle density. The wire would then cool to a lower temperature
relative to that of the thermometer. This cooling would then appear as an
enhancement of superconductivity. If this were the case, one could in
principle translate values of current into electron temperatures by relating
the resistive states with high currents at low temperatures to the resistive
states just above the critical temperature at low currents. As the former
resistive states can be destroyed by a weak magnetic field, one would expect
the same thing happen to the latter. However, as shown in Fig. 4, at low
currents, an applied magnetic field does not enhance superconductivity, but
destroys it above some critical value. This also distinguishes the present
observations from the negative magnetoresistance reported for Pb wires
\cite{Xiong}, which was attributed to fluctuations in the sign of the
Josephson coupling \cite{Kivelson}. In addition, the fact that magnetic field
affects the higher and lower resistance parts of the transition differently
provides support for the assertion that the effect is not thermal in origin.

A second possibility relates to the negative magnetoresistance of the
Cernox$^{\textregistered}$thermometer used in the PPMS $^{3}$He insert
\cite{Rosenbaum}. The response of the control system would be to interpret the
resistance change as a temperature increase. In order to maintain the set
point, the system would cool down, and the resistance of would be reduced
because the temperature is decreased. We rule this effect out by carrying out
an estimate of the magnetoresistance. For the resistance of the wire to drop
from its zero magnetic field value to zero at T = 0.6 K, the temperature would
have to fall below 0.46K. This would correspond to a magnetoresistance $\Delta
R/R$ $>$100\%, whereas the actual magnetoresistance is less than 10\% up to 2T
at 0.6K. This field is much higher than any in the experiment. Furthermore
there is a correction for magnetoresistance in the PPMS software, which
compensates for the magnetoresistance. As a consequence thermometer
magnetoresistance is irrelevant.

Finally, polarization of the magnetic moments of surface oxides, which would
quench pair-breaking spin fluctuations resulting in enhanced critical currents
\cite{Rogachev}, is also not relevant, as the field applied, tens of Oe, is
too weak to polarize impurity magnetic moments at 0.5K. In addition, one can
rule out compensation of the self-field by the applied field, since the
self-field, being the order of 0.1 Oe, is orders of magnitude smaller than the
applied field.%

\begin{figure}
[ptb]
\begin{center}
\includegraphics[
height=2.0038in,
width=2.399in
]%
{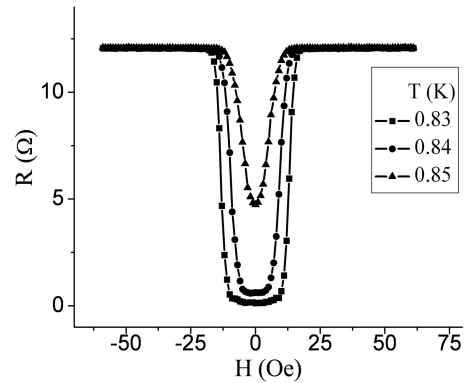}%
\caption{Magnetic field dependence of the wire resistance in the vicinity of
the transition temperature, at a low current with I = 0.4$\mu A$. }%
\end{center}
\end{figure}

We now suggest that the underlying physical mechanism for our observations is
damping of phase fluctuations by magnetic-field enhanced dissipation. From
numerous experiments on wires \cite{Newbower}, it is generally believed that
the resistive state at the bottom of the transition regime is characterized by
a nonzero order parameter with the resistance coming from phase slip processes
within the wire \cite{Little}. Frequently a resistively shunted Josephson
junction picture is used to describe the wire. In this picture the dynamics of
the phase are described by a tilted washboard potential and the applied
current determines the tilt \cite{Refael}. When the applied current is low,
the probability for phase slip is low at low temperature and is significant
only near T$_{c}$. As the current increases, approaching the depairing
current, the washboard potential is tilted further and the energy barrier for
phase slip is reduced \cite{Tinkham}. The wire is driven resistive when the
probability for a phase slip, and diffusion down the washboard potential
becomes detectable even at temperatures well below the transition temperature.
Dissipation provides damping for this process. This dissipation can originate
from the quasiparticles, either locally within the wire \cite{Refael,Zaikin1}
or in the electrodes connected to it \cite{Buchler}. For mesoscopic Josephson
tunneling junctions the phase can be localized in one of the wells of the
washboard potential when the shunt resistance falls below h/4e$^{2}$ resulting
in superconductivity \cite{Yamaguchi}. This is dissipation-induced
localization of the phase. The application of a magnetic field, even though
suppressing the superconductivity by smearing the density of states and
increasing the quasiparticle population and the quasiparticle conductance
channel, increases dissipation and therefore enhances the damping of phase
slip processes. This would appear to occur in a manner sufficient to localize
the phase, resulting in a return to the superconducting state. With further
increase of the magnetic field, too large a transport current or too high a
temperature, the barrier heights are no longer large enough to localize the
phase fluctuations and the resistive state is reentered.

Vodolazov \cite{Vodolazov1} has argued, based on a generalized time-dependent
Ginzburg-Landau equation \cite{Watts-Tobin}, that magnetic fields can enhance
superconductivity as a consequence of the magnetic field dependence of the
charge imbalance relaxation length and the presence of normal
metal/superconductor boundaries. In the present experiment, the fields at
which reentrance occurs are not sufficient to drive the electrodes into normal
state and temperatures are well below the transition temperature. As a
consequence we believe the considerations of Vodolzaov are not applicable.

Many of the samples we fabricated exhibited only very large negative
magnetoresistances and did not reenter the superconducting state. Examination
of these samples revealed that their transition temperatures and residual
conductivities were lower than those of samples which re-entered, suggesting
they were dirtier with shorter coherence lengths. SEM imaging showed that the
Zn electrodes of reentrant samples were smooth, and therefore the associated
wires should exhibit greater uniformity in their cross-section areas. If the
dissipation comes from the leads, the argument can be made that a shorter
coherence length results in weaker coupling between the wire and leads which
eventually results in a damping of fluctuations not sufficient to localize the
phase. It might also be argued that greater inhomogeneity in the wire
structure can produce more weak points in the wire, which remain in normal
state despite the damping of phase fluctuations.

In summary, we have demonstrated that applied current plays an important role
in driving phase fluctuations or phase slip in quasi-one dimensional
superconducting wires. When the current is close to the depairing value, the
wire enters a regime in which zero resistance is lost over a wide range of
temperatures well below the transition temperature. Unlike thermal phase slips
near T$_{c}$, this current driven phase slip regime can be damped and the
superconducting state reentered by enhancing the dissipation through
increasing the magnetic field. An important, perhaps unanswered question is
whether the resistive state results from thermal or quantum diffusion of the
phase. If the latter is the case, then the observed behavior, the reentrance
into the superconducting state with the application of a magnetic field, is an
example of a dissipative phase transition or the suppression of macroscopic
quantum tunneling of the phase by interaction with a dissipative environment
\cite{Leggett}.

This work was supported by the U. S. Department of Energy under grant DE-FG02-02ER46004.

\end{document}